\documentstyle[12pt]{article}
\begin{document}
\sloppy
\begin{center}
{\large \bf Batalin-Tyutin Quantisation of the $CP^{N-1}$ model}
\\[1cm]
 N. Banerjee\footnote{e-mail address:shila@saha.ernet.in}\\ Saha Institute of
 Nuclear Physics\\1/AF, Bidhannagar, Calcutta
700064\\
 India\\[0.5cm]
 Subir Ghosh\\Gobardanga Hindu college\\North 24-Parganas, West Bengal,
India\\[0.5cm]
                  and\\
R. Banerjee\\S. N. Bose National Center for Basic Sciences\\Sector I, DB 17,
Salt Lake\\
Calcutta- 700064, India.\\[1cm]
\end{center}
\begin{abstract}
The  $CP^{N-1}$  model  is  quantised  in the generalised canonical formalism
of
Batalin and Tyutin by converting the original second  class  system  into
first
class.  Operator  ordering  ambiguities present in the conventional
quantisation
scheme of Dirac are thereby avoided.
The  first  class  constraints,  the involutive Hamiltonian and the BRST
charge are explicitly  computed. The
partition function is defined and evaluated in the unitary gauge.
\end{abstract}
\newpage
\section{{\bf Introduction}}

The  quantisation  of  systems  with  second class constraints has always been
a
challenging problem. One possible way is to employ the pioneering work of
Dirac
[1].  Another  possibility  is  to  interpret  one  half  of  the  second
class
constraints as first  class,  and  consider  the  other  half  as  gauge
fixing
conditions  [2].  A  more recent approach is the generalised canonical
formalism
developed by Batalin and Fradkin [3], and, Batalin and Tyutin  (BT) [4].

The object of this paper is to discuss the quantisation of the $CP^{N-1}$
model
in the BT [4] formalism. It ought to be emphasised that the quantisation of
this
model (which is an example of a second class system) has been performed
earlier,
both  by using the method of Dirac [5] as well as by splitting the second class
constraints into
first class and gauge conditions [6,7]. In both these treatments, however,
there
are  unplesant  features.  The  Dirac  brackets,  as computed in [5], are
highly
nontrivial. They are field dependent and also have  a  nonpolynomial
structure.
Consequently  transition to the quantum theory is riddled with operator
ordering
ambiguities. In the other treatment [6,7],  on  the  other  hand,  there  is
an
ambiguity  in  the  splitting of constraints leading to a nonunique
Hamiltonian.
Moreover explicit computations [6,7] reveal that  the  modified  Hamiltonian
is
{\it nonlocal}.  All these bothersome features are bypassed in our analysis.
Following
BT [4], we convert  the  original  second  class  system  into  first  class
by
introducing  new  fields  in  an  extended phase space.The first class
constraints, the
unitarising Hamiltonian and the BRST  charge  are  explicitly  constructed.
The
phase  space  partition  function  is  defined. It is explicitly computed in
the
unitary gauge [3] by doing the relevant momentum integrals. The original
action
(corresponding  to the second class theory) is reproduced in this gauge, but
the
measure is nontrivial. The connection of our results for an analogus
computation
done by us [8] in the case of the $O(N)$ nonlinear sigma model is also
discussed.

We  may  mention  that  the basic idea of Batalin and Fradkin [3] or Batalin
and
Tyutin [4] of converting second class systems into first class has been
employed
previously  to quantise chiral gauge theories [9,10,11], the chiral boson
theory
[12], the massive Maxwell theory [10], the massive Yang-Mills  theory  [13]
and
the unitary gauge abelian Higgs model [14]. Most of these examples, however,
can
be simply quantised   by  the  conventional  formulation  of  Dirac  [1],
since
relevant  Dirac  brackets  are  simple  and  do  not  involve  operator
ordering
ambiguities. The same, as we stressed earlier,  is  not  true  in  the  case
of
$CP^{N-1}$  model.  Furthermore, the quantisation discussed in ref.[10] is not
a
systematic application of the Batalin-Fradkin [3] or Batalin-Tyutin [4]
method.
Our  analysis  clearly  demonstrates  the  necessity  of  a systematic
approach,
particularly when the algebra of constraints  is  field  dependent
as it is in the case of the $CP^{N-1}$ model.  Indeed  the
construction  of  the  involutive  Hamiltonian  requires  some involved
algebra.
Contrary to all previous computations [9--14], it contains an infinite number
of
terms. By proceding systematically we find a remarkable series of
cancellations
which  allow  us to express this infinite series as an exponential. We feel
that
the insights gained from our analysis would be useful in quantising other
types
of nonlinear models with a more involved constraint algrbra.

The  next  section discusses the quantisation of the model while our
conclusions
are presented in section 3.
\vspace{1cm}
\section{\bf Quantisation}

In order to implement the generalised canonical quantisation scheme of BT, it
is
necessary to specify the Hamiltonian together with its set of constraints.
This
is  done  by  first  considering  the  familiar  form  of  the Lagrangian of
the
$CP^{N-1}$ model,
\begin{equation}
{\cal L}~=~\partial_{\mu}z^{*}_{\alpha}\partial^{\mu}z^{\alpha}
-\frac{2g}{N}
(z_{\alpha}^{*}\partial^{\mu}z_{\alpha})(z_{\beta}\partial_{\mu}z_{\beta}^{*})
-\lambda(z_{\alpha}^{*}z_{\alpha} - \frac{N}{2g})
\end{equation}
 The canonical momenta are,
\begin{eqnarray}
\Pi_{\lambda} &=& \frac{\partial {\cal L}}{\partial\dot{\lambda}} =
0\nonumber\\
\Pi_{\alpha}  &=&   \frac{\partial    {\cal    L}}{\partial\dot{z}_{\alpha}}
=
M_{\alpha\beta} \dot{z}_{\beta}^{*}\nonumber\\
\Pi_{\alpha}^{*} &=& \frac{\partial {\cal L}}{\partial\dot{z}_{\alpha}^{*}}  =
M_{\alpha\beta}^{*} \dot{z}_{\beta}
\end{eqnarray}
 where,
\begin{equation}
M_{\alpha\beta}  =  \delta_{\alpha\beta}  - \frac{2g}{N}
z_{\alpha}^{*}z_{\beta}
\end{equation}
is a noninvertible matrix. The primary constraints of the theory are,
therefore,
given by
\begin{eqnarray}
T_{1} &=& \Pi_{\lambda}\approx 0\nonumber\\
T_{2} &=& \Pi_{\alpha}z_{\alpha} + \Pi_{\alpha}^{*}z_{\alpha}^{*}\approx
0\nonumber\\
T_{3} &=& \Pi_{\alpha}z_{\alpha} - \Pi_{\alpha}^{*}z_{\alpha}^{*}\approx 0
\end{eqnarray}
The canonical Hamiltonian is,
\begin{equation}
H_{C} =\int[ \Pi_{\alpha}^{*}\Pi_{\alpha} -
\partial_{i}z^{*}_{\alpha}\partial^{i}z^{\alpha}
+\frac{2g}{N}
(z_{\alpha}^{*}\partial^{i}z_{\alpha})(z_{\beta}\partial_{i}z_{\beta}^{*})
+\lambda(z_{\alpha}^{*}z_{\alpha} - \frac{N}{2g})]
\end{equation}
Time conserving the primary constraints leads to the familiar $CP^{N-1}$  model
constraint,
\begin{equation}
T_{4}= |z|^2 - \frac{N}{2g}
\end{equation}
and fixes $\lambda$ in (5) as,
\begin{equation}
\lambda=\frac{2g}{N}
[ \Pi_{\alpha}^{*}\Pi_{\alpha} +
\partial_{i}z^{*}_{\alpha}\partial^{i}z^{\alpha}
-\frac{4g}{N}
(z_{\alpha}^{*}\partial^{i}z_{\alpha})(z_{\beta}\partial_{i}z_{\beta}^{*})]
\end{equation}
No  further constraints are generated by this iterative sceme. Since the
$T_{1}$
constraint does not involve any dynamical variable,  we  henceforth  ignore
it.
Moreover inserting the value of $\lambda$ from (7) in (5), we find,
\begin{eqnarray}
\lefteqn{H_{C} =\int [(z^{*}_{\alpha} z_{\alpha})
(\Pi_{\alpha}^{*}\Pi_{\alpha})
-2 \partial_{i} z^{*}_{\alpha} \partial^{i}z^{\alpha}
(1-\frac{g}{N}|z|^{2})}\nonumber \\
 & & -\frac{2g}{N}
(z_{\alpha}^{*}\partial^{i}z_{\alpha})(z_{\beta}\partial_{i}z_{\beta}^{*})
(\frac{4g}{N}z_{\alpha}^{*}z_{\alpha} - 3)]
\end{eqnarray}
which, along with the constraints $T_{2}$, $T_{3}$, $T_{4}$, are the basic
inputs. The
Poisson algebra of constraints,
\begin{equation}
\{~T_{2}~,~T_{4}~\} = 2|z|^{2}\delta({\bf x}-{\bf y})
\end{equation}
with all other  brackets  being zero reveals that  $T_{2}$ and  $T_{4}$  are
second  class  constraints while $T_{3}$ is first class. We now convert the
second
class constraints into first class by  following  the  systematic  procedure
of
ref.[4].

The Poisson brackets (PB) among the second class constraints is compactly
expressed as,
\begin{eqnarray}
\lefteqn{\Delta_{ij}(x,y)= \big\{\Theta_{i}(x)~,~\Theta_{j}(y)\big\}}\nonumber
\\
& & = -2\epsilon_{ij}|z|^{2}\delta({\bf x}-{\bf y})
 ~~(i,j = 1,2) ~~~ (\epsilon^{12} =- \epsilon_{12} =1)
\end{eqnarray}
where we have made a change of notation,
\begin{equation}
\Theta_{1}(x) = T_{2}(x), ~~~\Theta_{2}(x) = T_{4}(x)
\end{equation}
The first class constraints $\Theta^{\prime}_{i}$ are then given by,
\begin{equation}
\Theta^{\prime}_{i}(z_{\alpha}, \Pi_{\alpha},\phi^{i})   =
\sum_{n=0}^{\infty}
{\Theta^{\prime}_{i}}^{(n)}, ~~~{\Theta^{\prime}_{i}}^{(n)} \sim (\phi)^{n}
\end{equation}
subject to the boundary condition,
\begin{equation}
{\Theta^{\prime}_{i}}^{(0)}   = \Theta^{\prime}_{i}(z_{\alpha},\Pi_{\alpha},
0)
= \Theta_{i}
\end{equation}
where $\phi^{i}$ are the new dynamical variables in the extended phase
space  $(z_{\alpha}, \Pi_{\alpha})\oplus  (\phi^{i})$ with the basic poisson
algebra [5],
\begin{equation}
\left\{ \phi^{i}(x)~,~\phi^{j}(y)\right\} = \omega^{ij}(x,y)
\end{equation}
and $\omega$ is an antisymmetric invertible matrix,
\begin{equation}
\omega^{ij}(x,y) = -\omega^{ji}(y,x).
\end{equation}
After (13), the next term in the series (12) is,
\begin{equation}
{\Theta^{\prime}_{i}}^{(1)} (x) = \int dy ~X_{ij}(x,y) \phi^{j}(y)
\end{equation}
where,
\begin{equation}
\int dz~dz^{\prime} \big[ X_{ij}(x,z) \omega^{jk}(z,z^{\prime}) X_{kl}
(z^{\prime},y)  \big] = - \Delta_{il}(x,y)
\end{equation}
with $\Delta_{il}(x,y)$ defined in (10).

An intelligent choice for $\omega^{ij}(x,y)$ and
$X_{ij}(x,y)$, which considerably simplifies the algebra, satisfying (15) and
(17) is
\begin{eqnarray}
\omega^{ij}(x,y) &= &2\epsilon^{ij}\delta(x-y) \nonumber\\
X_{ij} (x,y) &= &\left(
\begin{array}{lr}
 1 & 0 \\
0 & -|z|^{2}\\
\end{array}
\right) \delta(x-y).
\end{eqnarray}
Consequently,
\begin{equation}
\Theta_{1}^{\prime (1)} = \phi^{1},~~~\Theta_{2}^{\prime (1)} =
-|z|^{2}\phi^{2}
\end{equation}
The  other  terms  $(n>1)$ in the series (12), which are obtained by a
recursion
relation, vanish. Hence the first class constraints are,
\begin{eqnarray}
\Theta^{\prime}_{1} &= &\Theta_{1} +  \phi^{1} \nonumber\\
\Theta^{\prime}_{2} &= &\Theta_{2} - |z|^{2} \phi^{2}
\end{eqnarray}
which are strongly involutive,
\begin{equation}
\left\{ \Theta^{\prime}_{i}(x)~,~\Theta^{\prime}_{j}(y)\right\} = 0
\end{equation}

Having found the first class constraints, we now compute the corresponding
first
class Hamiltonian. This is given by,
\begin{equation}
H^{\prime} (z^{\alpha}, \Pi_{\alpha}, \phi^{i}) = \sum_{n=0}^{\infty} H^{\prime
(n)},
{}~~H^{\prime (n)} \sim (\phi)^{n}
\end{equation}
subject to the initial condition,
\begin{equation}
H^{\prime (0)} = H^{\prime} (z^{\alpha}, \Pi_{\alpha}, 0) = H_{C}
\end{equation}
The general expression for $H^{\prime (n)}$ is given in ref.[4],
\begin{equation}
H^{\prime (n+1)} = -\frac{1}{n+1}\int dx dy dz \left[\phi^{i}(x)
\omega_{ij}
(x,y) X^{jk}(y,z) G_{k}^{(n)}(z)\right] ~~(n\geq 0)
\end{equation}
where  $\omega_{ij}(x,y)$  and  $X^{jk}(y,z)$  are  the inverse
matrices  of  $\omega^{ij}(x,y)$   and    $X_{jk}(y,z)$ respectively,
defined  in  (18).  The generating functional $G^{(n)}_{k}$ has a
very simple form,
\begin{eqnarray}
G_{k}^{(0)} &= &\left\{\Theta_{k}~,~H_{C}\right\}\nonumber\\
G_{k}^{(n)} &= &\left\{\Theta_{k}^{\prime (1)}~,~H^{\prime
(n-1)}\right\}_{(z_{\alpha},\Pi_{\alpha})}
+ \left\{\Theta_{k}~,~H^{\prime
(n)}\right\}_{(z_{\alpha},\Pi_{\alpha})}~~(n\geq 1)
\end{eqnarray}
which is a consequence of the judicious  choice   (18)  so  that the
series  (13) comprises only two terms $\Theta_{i}$ and $\Theta_{i}^{\prime
(1)}$.
The remarkable algebraic simplification achieved in (25) can be appreciated by
looking at the general structure for $G_{k}^{(n)}$ given in equation
(2.54)   of   [4]. The  symbol  $\{~,~\}_{(z_{\alpha}, \Pi_{\alpha})}$
appearing in (25) means that the relevant PB has to be computed with
respect to those variables.  Using (18)to (25)
all the terms in the series  (24) may be evaluated. Interestingly, in contrast
to  (12), it turns out to be an infinite series. We find, however,
that a remarkable sequence of cancellations occurs leading to the result,
\begin{equation}
H^{\prime}   =   H_{C}   -   \frac{g}{N}\int   dx   \phi^{2}
|z|^{2}\Theta_{2}
+\frac{g}{2N}  \int  dx \phi^{2}\phi^{2} (|z|^{2})^{2} +
\sum_{p=1}^{\infty}H^{(p)}
\end{equation}
where,
\begin{eqnarray}
H^{(p)} = \int dx_{1} dx_{2}\ldots dx_{p}& [\frac{(-1)^{p}}{p!} \frac{1}{2}
(\frac{\phi^{1}}{|z|^{2}}) (x_{1})\{ \Theta_{2}(x_{1})~,~
\frac{1}{2} (\frac{\phi^{1}}{|z|^{2}}) (x_{2})\{ \Theta_{2}(x_{2})\ldots
\nonumber\\
 &\frac{1}{2}  (\frac{\phi^{1}}{|z|^{2}})
(x_{p})\{\Theta_{2}(x_{p})~,~H_{0}\}\}\}_{p-fold}]
\end{eqnarray}
and,
\begin{equation}
H_{0} = H_{C} - \frac{2g}{N}\int dx~ |z|^{2}\Pi^{*} \Pi
\end{equation}
is a function of $z_{\alpha}$ fields only. A  compact  way  to  express  (27)
is,
to use the functional Schr\"odinger representation $ (\Pi_{\alpha}\to (-)
\frac{\delta}{\delta z_{\alpha}})$ so that,
\begin{equation}
H^{(p)} =  \frac{1}{p!}\int  dx_{1}\ldots  dx_{p}
\left[\frac{\phi^{1}}{2|z|^{2}}
\left\{z_{\alpha}\frac{\vec{\delta}_{L}}{\delta z_{\alpha}}
+z_{\alpha}^{*}\frac{\vec{\delta}_{L}}{\delta z_{\alpha}^{*}}
\right\}\right]^{p} H_{0}
\end{equation}
where $\vec{\delta_{L}}$ is the left derivative. Combining (29) with (26)
yields the final Hamiltonian,
\begin{eqnarray}
H^{\prime} =\frac{2g}{N} \int |z|^{2}|\Pi|^{2}  &-& \frac{g}{N} \int
\phi^{2}|z|^{2}\Theta_{2}
+ \frac{g}{2N} \int \phi^{2}\phi^{2} (|z|^{2})^{2}\nonumber\\
&+&\int \exp \left[\int \frac{\phi^{1}}{2|z|^{2}}
\left\{z_{\alpha}\frac{\vec{\delta}_{L}}{\delta z_{\alpha}}
+z_{\alpha}^{*}\frac{\vec{\delta}_{L}}{\delta z_{\alpha}^{*}}
\right\}\right]H_{0}
\end{eqnarray}
which is  strongly  involutive  with  the constraints $\Theta^{\prime}_{i}$,
\begin{equation}
\left\{ H^{\prime}~,~ \Theta^{\prime}_{i}\right\} = 0
\end{equation}
We have thus converted  the second class system (with
constraints $\Theta_{1}$, $\Theta_{2}$ and Hamiltonian
$H_{C}$  )   into   first   class   (with  constraints   $\Theta^{\prime}_{1}$,
$\Theta^{\prime}_{2}$ and Hamiltonian
$H^{\prime}$).

Now  we have to include the original first class constraint $T_{3}$ (4) into
our
analysis. It is simple to  see  that  the  first  class  nature  of  $T_{3}$
is
preserved   with   respect   to   the   new  constraints
$\Theta_{1}^{\prime}$,
$\Theta^{\prime}_{2}$,
\begin{equation}
\big\{T_{3}~,~\Theta^{\prime}_{i}\big\} = 0
\end{equation}
which is a consequence of the fact that the new constraints
$\Theta_{i}^{\prime}
$  differ  from the old one $\Theta_{i}$ by fields $(\phi^{1}, ~\phi^{2})$
which
have  vanishing  PB  with  $T_{3}$.  It  is  also  found  that  new
Hamiltonian
$H^{\prime}$  is  in  involution  with  $T_{3}$. This is not obvious, but can
be
verified by an explicit calculation,
\begin{equation}
\{ H^{\prime}~,~T_{3}\} =
-\frac{8g}{N}\partial^{i}[(1+\frac{\phi^{1}}{|z|^{2}})
(z^{*}\stackrel{\leftrightarrow}{\partial_{i}}z)\Theta_{1}^{\prime}
\end{equation}
This  completes  the  operatorial conversion of the original second class
theory
into first class. Making a change of notation $T_{3}\to\Theta^{\prime}_{3}$,
we
can  combine  (21)  and  (32)  as  well  as (31) and (33), to write the
complete
involution algebra as follows:
\begin{eqnarray}
\big\{ \Theta_{\alpha}^{\prime}~,~\Theta_{\beta}^{\prime}\big\} = 0 ,~~~\alpha,
\beta = 1,2,3\\
\big\{H^{\prime}~,~\Theta_{\alpha}^{\prime}\big\} = \int  dy
V^{\beta}_{\alpha}
(x,y)\Theta^{\prime}_{\beta} (y)
\end{eqnarray}
where,
\begin{equation}
V^{\beta}_{\alpha}        =       -[\frac{8g}{N}
(1+\frac{\phi^1}{|z|^{2}}
(z^{*}\stackrel{\leftrightarrow}{\partial_{i}}z)]   \delta_{\alpha
1}\delta_{\beta    3}
\partial^{i}_{x}\delta({\bf x} - {\bf y})
\end{equation}
The above equations clearly illuminate the first class nature of the system.

We next construct the BRST invariant Hamiltonian $H_{BRST}$ and charge $Q$,
which are given by,
\begin{equation}
 H_{BRST}= H^{\prime} + \int dx P^{3}(x) V^{1}_{3} (x,y) \bar{C}_{1} (y)
\end{equation}
\begin{equation}
Q = \int dx \left[C^{\alpha}(x)\Theta^{\prime}_{\alpha}(x) + p_{\alpha}(x)
P^{\alpha}
(x)\right];~~ \alpha = 1,2,3
\end{equation}
where $(C^{\alpha}~,~\bar{P}_{\beta})$ and $(P^{\alpha}~,~\bar{C}_{\beta})$
form
a  pair of canonical  ghost (antighost) having the opposite Grassman parity as
$\Theta_{\alpha}^{\prime}$;
\begin{equation}
\left\{ C^{\alpha}(x)~,~\bar{P}_{\beta}(y)\right\} = \left\{ P^{\alpha}(x)~,~
\bar{C}_{\beta}(y)\right\}  =  \delta^{\alpha}_{\hphantom{\alpha}\beta}
\delta ({\bf x}-{\bf y})
\end{equation}
while  $(p_{\alpha}~,~q^{\beta})$  is a canonical multiplier set with the same
Grassman parity as $\Theta_{\alpha}$,
\begin{equation}
\left\{ q^{\alpha}(x)~,~{p}_{\beta}(y)\right\}
= \delta^{\alpha}_{\hphantom{\alpha}\beta}   \delta (x-y)
\end{equation}
Finally, the physical Hilbert space is defined by,
\begin{equation}
Q|phys\rangle = 0, ~~~~~~~~|phys\rangle \neq  Q|...\rangle
\end{equation}
This   completes   the   operator   formulation   of  the model. We now turn
our
attention to the partition function. Let us  first define  the  gauge  fermion
operator
$\psi$ given in ref.[3]
\begin{equation}
\psi  = \int dx \left[\bar{P}_{\alpha}q^{\alpha} +
\bar{C}_{\alpha}\chi^{\alpha}
\right]
\end{equation}
where  $\chi_{\alpha}$ is the hermitean gauge fixing function with identical
Grassman parity as $\Theta_{\alpha}$ and satisfy,
\begin{equation}
det|\{ \chi_{\alpha}~,~\Theta^{\prime}_{\beta}\}|\neq 0
\end{equation}
The complete unitarising Hamiltonian $H_{U}$ is now defined by,
\begin{equation}
H_{U} = H_{BRST} + \{\psi~,~Q\}
\end{equation}
We next rename the variables $\phi^{1}$ and $\phi^{2}$ as,
\begin{equation}
\phi^{1}\to 2\phi,~~~~~~~~~\phi^{2}\to \Pi_{\phi}
\end{equation}
so  that $(\Pi_{\phi}~,~\phi)$ may be considered as a canonically conjugate
pair by
virtue of (14) and (18). Then the  partition function $Z$ is given by,
\begin{equation}
Z = \int [d\mu] e^{iS}
\end{equation}
where,
\begin{equation}
S =\int \left[\Pi_{\alpha}\dot{z}_{\alpha} +
\Pi^{*}_{\alpha}\dot{z}^{*}_{\alpha}+ \Pi_{\phi}\dot{\phi} +
C^{\alpha}\dot{\bar{P}}_{\alpha} + P^{\alpha}\dot{\bar{C}}_{\alpha}  +
p_{\alpha}
\dot{q}^{\alpha} - H_{U}\right]
\end{equation}
where the measure $[d\mu]$ includes all the variables appearing in the action.
Different  choices  for  the  gauge  function  $\psi$  can be done to
explicitly
evaluate the partition function. The final result for  $Z$  is,  however,
gauge
independent by the Fradkin-Vilkovisky theorem [15,16]

Before concluding this section we shall work out $Z$ in the `unitary  gauge'
[3,5,],
which amounts to choosing two of the gauge conditions to be the original second
class constraints,
\begin{equation}
\chi_{1} = \Theta_{1},~~\chi_{2} = \Theta_{2},~~\chi_{3}=z_{1} + z^{*}_{1}
\end{equation}
Making  the   change   of   variables   $\chi_{\alpha}\to \chi_{\alpha}/\beta$,
$p_{\alpha}  \to  \beta p_{\alpha}$, $\bar{C}_{\alpha}\to
\beta\bar{C}_{\alpha}$
whose (super) Jacobian is unity , and then taking the limit $\beta\to 0$ [16],
we obtain,
\begin{eqnarray}
\lefteqn{Z = \int \left[{\cal D}z_{\alpha}  {\cal D} \Pi_{\alpha}{\cal
D}z^{*}_{\alpha}
{\cal D}\Pi^{*}_{\alpha}
{\cal D}\phi {\cal D}\Pi_{\phi}
\right]}\nonumber\\
& &\delta (\Theta_{1}) \delta(\Theta_{2})\delta(z_1+z^{*}_{1})
 \delta(\phi)   \delta(\Pi_{\phi})\delta(\Theta_{3}) det |z_{1}| e^{iS}
\end{eqnarray}
with,
\begin{equation}
S = \int\left(\Pi_{\alpha}\dot{z}^{\alpha} +
\Pi_{\alpha}^{*}\dot{z}_{\alpha}^{*}
+ \Pi_{\phi}\dot{\phi} - H_{C}\right)
\end{equation}
We next perform the momentum integrations. The $\Pi_{\phi}$ integral is
trivial.
The delta functions involving the original constraints are  expressed  by
their
corresponding  Fourier  transforms.  Then  the  integrals  over
$\Pi_{\alpha}$,
$\Pi^{*}_{\alpha}$ are done. This yields, for the action,
\begin{eqnarray}
S = \partial^{\mu}z^{*}_{\alpha}\partial_{\mu}z_{\alpha} - \frac{2g}{N}(z^{*}
\partial^{i}z) (z\partial_{i}z^{*}) &+&\frac{N}{2g} \xi^{2}\nonumber\\
&+& 2\xi \dot{z}z^{*} + \frac{N}{2g}\eta^{2}
\end{eqnarray}
where  $\xi$ and $\eta$ are  the  Fourier variables. The integral over $\eta$
is a trivial Gaussian. Finally the $\xi$ integration is done
to yield,
\begin{equation}
S = \partial^{\mu}z^{*}_{\alpha}\partial_{\mu}z_{\alpha} - \frac{2g}{N}(z^{*}
\partial^{\mu}z) (z\partial_{\mu}z^{*})
\end{equation}
while the measure is,
\begin{equation}
[d\mu] = \delta(z_{1}+z_{1}^{*})\delta(|z|^{2}-\frac{N}{2g}) det|z_{1}|{\cal
D}z_{\alpha}
{\cal D}z_{\alpha}^{*}
\end{equation}
The expression for the original (classical) action (1) is seen to be
reproduced.
The  term
involving  the Lagrange  multiplier  $\lambda$ in (1) is manifested here
through the
delta function $\delta(|z|^{2} -\frac{N}{2g})$  appearing in  the  measure
(49).
Other  forms  of  action  with  corresponding  measures can be obtained by
other
choices of gauges.

\vspace{1cm}
\section{\bf Conclusion}

We have systematically applied the  generalised  canonical  formalism  [3,4]
to
quantise the $CP^{N-1}$ model, which is an example of a second class theory.
The
quantisation of  second class systems, it may be recalled, is  usually  done
by
the method of Dirac [1].
In  this  particular  case,  however,  the  Dirac  brackets  are known [5] to
be
extremely complicated. They  are  field  dependent  and  have  a
nonpolynomical
structure.  Consequently transition to the quantum theory is plagued with
severe
operator ordering ambiguities. All these problems are bypassed  here  since
the
brackets  are canonical. Moreover by converting the original second class
system
  into first  class  in  an  extended  phase  space,  the  quantisation
program
simplifies  since  there already exists a well established generalised
canonical
scheme [15,16] for quantising such systems.

It is worthwhile  to  point  out  the  difference  of  our  results  with
other
(non-Dirac)  approaches, notably those presented in ref.[6,7]. In these papers
the
pair of second class constraints is interpreted as  a  combination  of  a
first
class  constraint  and a gauge fixing constraint. The hamiltonian,
consequently,
gets modified. In fact, it turns  out  to  be  {\it  nonlocal}.  Our
involutive
Hamiltonian,  on  the  contrary,  is  local.  Moreover  there is an
arbitrariness in the decomposition of the second class constraints into a
first
class  constraint  and a gauge constraint, so that the corresponding
Hamiltonian
is not unique.  The  involutive  Hamiltonian  in  our  treatment  is
determined
uniquely  once  the  matrices  $\omega_{ij}$ and $X^{ij}$ (18) have been
chosen.
Although   there   is   a   `natural   arbitrariness'   in   the    choice
of  these  matrices,  these  correspond  to  canonical transformation [4] in
the
extended  phase  space.  Thus  two  involutive  Hamiltonians  resulting  from
a
different choice of matrices $\omega_{ij}$, $X^{ij}$ are canonically
equivalent.

The  general  philosophy  of  Batalin-Fradkin  [3]  and  Batalin-Tyutin  [4]
of
converting  second  class systems into forst class ones has been used
previously
to discuss the quantisation of Proca model [10] and the chiral  Schwinger
model
[9,10].  However these  analyses  are  unsystematic and could be carried
through principally
because the algebra of constraints was very simple. In the present example
where
this  algebra  becomes field dependent, such an approach would be untenable.
The
construction  of  the  involutive  Hamiltonian  drives  home  this  point.
This
Hamiltonian   comprises  an infinite number  of  terms  and  only  a
systematic  series  of
cancellations  (whose  origin  is  contained  in  the  intelligent   choice
of
$\omega^{ij}$  and  $X_{ij}$  )  could  enable  us  to  express  it  as a
closed
(exponential) form.

Finally, we would like to make a connection of our result  with  those
obtained
previously  by  us  [8]  concerning  the $O(N)$ invariant nonlinear sigma
model.
Following the BT prescription we had shown there that the partition function
in
the unitary gauge had the form,
\begin{equation}
Z   =\int  dn^{a}  \delta(n^{2}-1)  \exp[i\int  \frac{1}{4}\partial^{\mu}
n^{a}
\partial_{\mu}n^{a}]
\end{equation}
where $n^{a}$ are the sigma model fields constrained by the delta function.
For
the  special  case  of  the $O(3)$ sigma model and $CP^{1}$ model it can be
seen
from (49) and (53) that the action in two cases becomes identical, on using
Hopf
map
\begin{equation}
n^{a} = z^{*}\sigma^{a} z, ~~\mbox{($\sigma$ are Pauli matrices)}
\end{equation}
but   the   measure  does  not  agree.  Contrary  to  the  $O(3)$  example,
the
Faddeev-Popov determinant (50) for the $CP^{1}$ model is found to be
nontrivial.

\end{document}